# Low-Dose CT Imaging Using a Regularization-Enhanced Efficient Diffusion Probabilistic Model

Qiang Li[1], Mojtaba Safari[1], Shansong Wang[1], Huiqiao Xie[1], Jie Ding[1],

Tonghe Wang[2] and Xiaofeng Yang[1]*

[1]Department of Radiation Oncology and Winship Cancer Institute, Emory University, Atlanta, GA 30322

[2]Department of Medical Physics, Memorial Sloan Kettering Cancer Center, New York, NY 10065

Correspondence to: xiaofeng.yang@emory.edu

# Abstract

**Background:** Low-dose CT (LDCT) imaging reduces patient radiation exposure but introduces elevated noise levels that degrade image quality and undermine downstream clinical tasks such as diagnosis and quantitative analysis. Existing denoising approaches often require extensive diffusion steps that impede real-time clinical applicability.

**Purpose:** To address this challenge, we propose a regularization-enhanced efficient diffusion probabilistic model (RE-EDPM), a rapid and high-fidelity denoising framework that incorporates residual guidance between low dose and full dose CT scans and employs hybrid perceptual and total variation regularization to preserve anatomical fidelity and diagnostic quality.

**Methods:** RE-EDPM incorporates a residual-shifting mechanism into the forward diffusion process to better align the LDCT and FDCT distributions, followed by four reverse diffusion steps using a Swin-based U-Net backbone. A composite loss function combining pixel-level reconstruction, perceptual similarity (LPIPS), and spatially total variation (TV) is used to suppress spatially varying noise while preserving fine structural details. We evaluated RE-EDPM on a public LDCT benchmark dataset across different dose levels and anatomical sites. Quantitative performance was assessed using Structural Similarity Index Measure (SSIM), Peak Signal-to-Noise Ratio (PSNR), and Visual Information Fidelity (VIFp), and RE-EDPM was compared against several state-of-the-art methods.

**Results:** On public LDCT benchmarks at 10% dose for chest and 25% dose for abdomen, RE-EDPM achieved SSIM of $0.879 \pm 0.068$, PSNR of $31.60 \pm 2.52$ dB, and VIFp of $0.366 \pm 0.121$ for chest images, and SSIM of $0.971 \pm 0.000$, PSNR of $36.69 \pm 2.54$ dB, and VIFp of $0.510 \pm 0.007$ for abdominal images. Visualizations of residual and difference maps confirmed superior noise suppression and structural fidelity. Ablation studies and Wilcoxon signed-rank tests ($p < 0.05$) validated the significant contributions of the residual-shifting mechanism and each regularization component. RE-EDPM processes two $512 \times 512$ slices in approximately $0.25$ s ($\approx 0.125$ s per slice) on modern high-performance hardware, supporting near-real-time clinical application.

**Conclusions:** RE-EDPM offers robust LDCT denoising with minimal inference time, achieving an optimal balance between noise reduction and anatomical preservation. Its efficiency and high performance make it a strong candidate for real-time clinical deployment and for broader applications in transforming low-quality to high-quality medical imaging tasks.

**Keywords** Low-dose CT; denoising; diffusion probabilistic model; residual-shifting; Swin U-Net; perceptual regularization; learned perceptual image patch similarity; total variation loss.

# 1. Introduction

Computed tomography (CT) is one of the most widely used imaging modalities in clinical practice, offering high-resolution visualization of anatomical structures(Buzug 2011). However, the ionizing radiation associated with CT scans poses potential health risks, including an increased likelihood of malignancies and metabolic abnormalities(Albert 2013, Smith-Bindman *et al* 2025). To minimize radiation exposure, low-dose CT (LDCT) has been adopted as an alternative to standard-dose CT. While LDCT reduces the radiation burden on patients, it introduces significant noise and artifacts, which can degrade image quality and impair diagnostic accuracy. CT noise originates from both electronic and quantum sources, typically modeled as Gaussian and Poisson distributions, respectively(Wang *et al* 2008, Kalra *et al* 2004). These characteristics make LDCT image enhancement a challenging yet essential task for maintaining diagnostic integrity.

A wide spectrum of denoising approaches has been explored. Classical techniques include sinogram-domain filtering, iterative reconstruction, and image-domain post-processing(Willemink and Noël 2019, Willemink *et al* 2013, Kiss *et al* 2024). Sinogram filtering suppresses noise prior to reconstruction but suffers from resolution loss and vendor dependence. Iterative reconstruction incorporates prior knowledge to improve image quality but is computationally intensive and less scalable(Sidky and Pan 2008, Ziegler *et al* 2007, Sidky *et al* 2006, Ramani and Fessler 2011). Image-domain methods such as non-local means, dictionary learning(Xie *et al* 2018, Ma *et al* 2011, Chen *et al* 2013) and BM3D offer better efficiency but may over smooth anatomical structures or require parameter tuning due to their limited adaptability to CT-specific noise(Feruglio *et al* 2010). Recent advancements in DL have enabled data-driven denoising by learning mappings between LDCT and full dose CT (FDCT) pairs(Diwakar and Singh 2020, Singh *et al* 2022, Diwakar *et al* 2023). RED-CNN perceptual-loss-based networks(Chen *et al* 2017b), GANs(Yang *et al* 2018),(Wang *et al* 2019), attention-enhanced residual models(Wu *et al* 2024), and back-projection-based frameworks(Shu and Entezari 2024) have shown significant improvements. However, many still struggle with preserving subtle lesion features, and excessive smoothing may compromise diagnostic accuracy.

Denoising Diffusion probabilistic models (DDPMs) have recently emerged as a promising generative approach for medical image processing (Safari *et al* 2024, 2025, n.d.). DDPMs progressively corrupt images with noise in a forward process and learn to invert this process through iterative denoising. While effective, their clinical utility is limited by the heavy computational cost: hundreds to thousands of reverse steps are typically required to achieve high-quality reconstructions. Moreover, conventional U-Net-based noise prediction models used in DDPMs may inadequately capture the complex texture details of CT. We introduce a residual error shifting mechanism, which steers the forward process along the residual between LDCT and FDCT, ensuring that the reverse dynamics remain closely conditioned on the measurement. Second, we employ a noise-shifted initialization, starting the reverse process from a small stochastic neighborhood around the degraded image. This keeps the reverse dynamics well-conditioned for few-step samplers and enables a scheduler that allocates most of the limited budget to early high-frequency denoising, followed by rapid convergence to a high-fidelity solution. In addition, to account for the spatially variant nature of CT noise, we design a hybrid regularization scheme. MSE loss is retained as the primary objective to maintain theoretical consistency with KL minimization. Complementary terms include MAE (robust to outliers), a spatially total variation (TV) loss (promoting local smoothness while preserving edges), and learned perceptual image patch similarity (LPIPS) loss(Zhang *et al* 2018b) (encouraging perceptual similarity in a deep feature space). This combination jointly enforces pixel-wise accuracy, structural preservation, and perceptual quality.

This design allows our framework to achieve high-fidelity LDCT reconstruction with drastically fewer reverse steps, greatly reducing computational cost compared with conventional DDPMs. Whereas a vanilla DDPM requires ~1000 denoising steps and 15–30 s per 512×512 slice(Su *et al* 2025, Xia *et al* 2022, Liu *et al* 2025, Pan *et al* 2023). Our network denoises two slices in around 0.25 s on an NVIDIA A6000. This means a full thoracic CT scan of ~300 slices can be processed in under 40 s, rather than the >90 min typical of vanilla DDPMs, enabling true near real-time integration into clinical workflows. We introduce RE-EDPM (Regularization-Enhanced Efficient Diffusion Probabilistic Model), a novel framework for LDCT denoising. The main contributions are:

(i) Residual-shifting diffusion process: accelerates convergence and reduces artifacts, achieving high-quality LDCT denoising in only four reverse steps.

(ii) Hybrid adaptive regularization: integrates TV, MAE, and LPIPS with MSE to address spatially variant noise while preserving diagnostically important details.

(iii) Efficiency advantage: compared with standard DDPMs, RE-EDPM requires ~10× fewer training resources and achieves up to 250× faster inference. For example, a thoracic CT scan of ~300 slices can be reconstructed in under 40 s, compared to >90 min with vanilla DDPMs.

(iv) Comprehensive validation: extensive experiments on chest and abdomen datasets demonstrate that RE-EDPM consistently outperforms CNN-, GAN-, and diffusion-based baselines, highlighting its robustness and clinical potential.

## 2 Methods and data

## 2.1 Dataset

For our benchmark setup, we utilize a subset of the LDCT and Projection Dataset(Moen *et al* 2021), which comprises 200 clinical cases collected from two scanner vendors: 100 cases from GE Healthcare (Discovery CT750i) and 100 cases from Siemens Healthineers (SOMATOM Definition AS+ and SOMATOM Definition Flash). Each case includes both routine-dose and simulated low-dose projections, generated via noise insertion in the projection domain. From this dataset, we select 50 abdomen and 50 chest cases from each of the two scanner vendors, yielding 100 abdomen and 100 chest volumes in total. Simulated low-dose reconstructions are provided at 25% dose for the abdomen and 10% dose for the chest. The data are split per modality into 70% training, 20% validation, and 10% test sets. All images are linearly normalized to zero mean and unit variance.

During evaluation, testing was restricted to anatomically relevant axial regions: lung-containing slices for chest CT and non-lung slices for abdominal CT. All data partitioning and preprocessing steps were standardized and are fully reproducible, with the complete implementation released in our benchmark suite.

## 2.2 Proposed method

We propose a regularization-enhanced efficient diffusion probabilistic model (RE-EDPM) for LDCT reconstruction. The goal of RE-EDPM is to recover FDCT images $x^{FD}$ from their corresponding low-dose inputs $x^{LD}$. Following the DDPM framework, the model formulates a forward degradation process and a reverse restoration process. Unlike conventional DDPMs that diffuse images toward a standard Gaussian prior $\mathcal{N}(0, I)$, RE-EDPM introduces a residual-shift strategy: the forward process progressively shifts $x^{FD}$ toward the residual neighborhood of its paired low-dose image $x^{LD}$. During the forward diffusion process, we progressively add Gaussian noise to $x^{FD}$, obtaining a noisy version $x_t^{FD}$ at time step $t$. Specifically, the degradation endpoint is defined as $x_T^{FD} \sim \mathcal{N}(x^{LD}, \gamma^2 I)$, guided by the residual error $e_0 = x^{LD} - x^{FD}$. This design aligns the forward process with the measurement distribution, thereby shortening the diffusion trajectory required during reconstruction (illustrated in **Figure 1**). The reconstruction network in RE-EDPM adopts a U-Net backbone augmented with Swin Transformer blocks, similar in spirit to recent diffusion-based restoration studies(Yue *et al* 2024, Safari *et al* 2025, Ho *et al* 2020a). The use of Swin Transformers enlarges the receptive field and improves robustness to variations in image resolution enabling the model to better capture long-range contextual dependencies(Yue *et al* 2024). The resulting architecture (**Figure S1**) consists of convolutional layers, residual blocks, hierarchical downsampling and upsampling, and strategically embedded Swin Transformer modules to enhance structural fidelity and preserve fine details. Formally, the RE-EDPM model takes three inputs: (i) a low-dose CT image $x^{LD}$, (ii) its diffused representation $x_t^{FD}$ at diffusion step $t$, and (iii) the timestep $t$. The network learns to predict the clean full dose image $\hat{x}_t^{FD}$ from above inputs, formulated as $\hat{x}_t^{FD} = g_\varphi(x_t^{FD}, x^{LD}, t)$. The output $\hat{x}_t^{FD}$ serves as the denoised estimate of the ground-truth full-dose target $x^{FD}$. Training is conducted with a composite loss function combining MSE, MAE, LPIPS, and spatially adaptive TV regularization, jointly optimizing for pixel-wise accuracy, perceptual fidelity, structural sharpness, and smoothness.

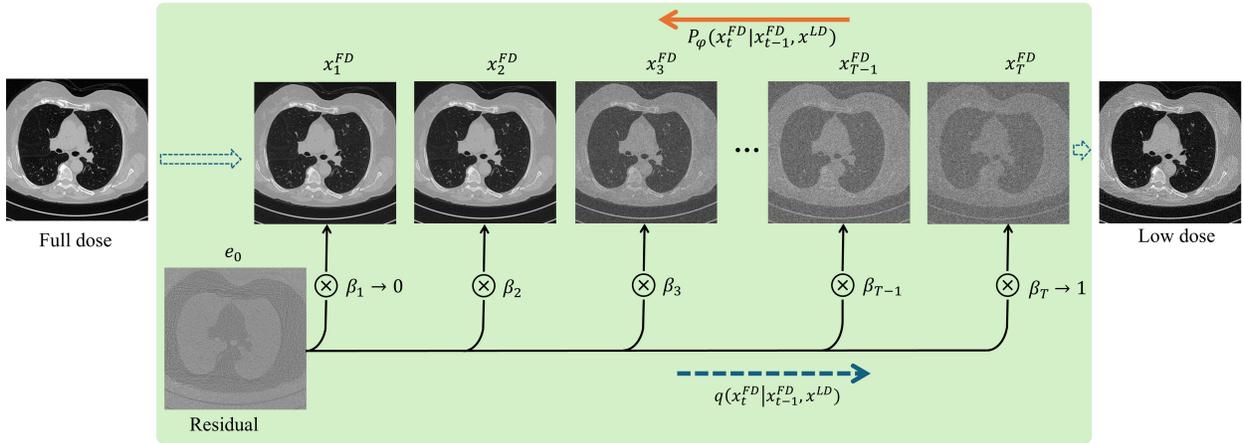

**Figure 1**: Overview of the proposed RE-EDPM. Our method builds up a Markov chain between the Full /Low dose CT image pair by shifting their residuals.

**Forward process.** To simulate the forward diffusion process, a monotonically increasing shifting sequence $\{\beta\}_{t=1}^T$ over time steps $t$ with bounding conditions $\beta_1 \to 0$ and $\beta_T \to 1$ is used. The transition kernel for simulating the forward diffusion process is given in (1), which is constructed based on the Markov chain and the residual error $e_0$ shift sequences (see Figure 1):

$$q(x_t^{FD}|x_{t-1}^{FD}, x^{LD}) = \mathcal{N}(x_t^{FD}; x_{t-1}^{FD} + e_0\alpha_t, \gamma^2\alpha_t I), \quad t \in [1, T] \qquad (1)$$

where $\alpha_1 = \beta_1 \to 0$ and $\alpha_t = \beta_t - \beta_{t-1}$ for $t > 1$, and $\gamma$ is a hyper-parameter introduced to improve the flexibility of the forward diffusion process. By framing the process as a Markov chain, the image at time step t can be generated from the image at step t–1 via the reparameterization trick, as follows: $x_t = x_{t-1} + \alpha_t e_0 + \sqrt{\gamma^2 \alpha_t} \epsilon_t$. However, iterating this update for every intermediate step is computationally expensive. Starting from $x_t$, we unroll the recursion back to $x_0$:

$$x_t = x_0 + \sum_{i=1}^{t} \left(\alpha_i e_0 + \sqrt{\gamma^2 \alpha_i} \epsilon_i\right)$$

$$= x_0 + \left(\sum_{i=1}^{t} \alpha_i\right) e_0 + \sum_{i=1}^{t} \sqrt{\gamma^2 \alpha_i}\, \epsilon_i. \quad (2)$$

Noting that $\sum_{i=1}^{t} \sqrt{\gamma^2 \alpha_i}\, \epsilon_i$ is itself Gaussian with variance $\gamma^2 \sum_{i=1}^{t} \alpha_i$, we can write the closed-form as: $x_t = x_0 + A_t e_0 + \sqrt{\gamma^2 A_t}\, \epsilon$, where $A_t = \sum_{i=1}^{t} \alpha_i$, $\epsilon \sim \mathcal{N}(0, I)$.

Here, we omit the superscript FD for brevity. The second term (mean) and the square of the third term (variance) in the summation given are equal to $\beta_t$. Thus, the marginal distribution at any time step t can be computed analytically as follows:

$$q(x_t^{\text{FD}} \mid x^{\text{FD}}, x^{\text{LD}}, t) = \mathcal{N}(x_t^{\text{FD}}; x^{\text{FD}} + e_0 \beta_t, \gamma^2 \beta_t I), \quad t \in [1, T] \quad (3)$$

**Reverse process.** This process trains a model, that employs a U-net backbone in which the conventional attention layers are replaced by Swin Transformer blocks(Su *et al* 2025) to improve generalization across different image resolutions(Xia *et al* 2022). To recover a FDCT image $x^{\text{FD}}$ from its low-dose counterpart $x^{\text{LD}}$, we model the posterior:

$$p(x^{\text{FD}} \mid x^{\text{LD}}) = \int p(x_T^{\text{FD}} \mid x^{\text{LD}}) \prod_{t=1}^{T} p_\varphi(x_{t-1}^{\text{FD}} \mid x_t^{\text{FD}}, x^{\text{LD}})\, dx_{1:T}, \quad (4)$$

where $p(x_T^{\text{FD}} \mid x^{\text{LD}}) \approx \mathcal{N}(x^{\text{FD}}; x^{\text{LD}}, \gamma^2 I)$ and $p_\varphi(x_{t-1}^{\text{FD}} \mid x_t^{\text{FD}}, x^{\text{LD}})$ is a reverse transition kernel that aims to learn $x_{t-1}^{\text{FD}}$ from $x_t^{\text{FD}}$ by training a network $g_\varphi$. Similar to conventional diffusion models (Song *et al* 2020b, Luo 2022, Ho *et al* 2020a), and each reverse transition can be written as follows by adopting the Gaussian assumption:

$$p_\varphi(x_{t-1}^{\text{FD}} \mid x_t^{\text{FD}}, x^{\text{LD}}) = \mathcal{N}\left(x_{t-1}^{\text{FD}}; \mu_\varphi(x_t^{\text{FD}}, x^{\text{LD}}, t), \Sigma_\varphi(t)\right) \quad (5)$$

The parameters $\varphi$ are found by minimizing the sum of KL divergences between the true forward kernel $q$ and our learned reverse kernel $p_\varphi$:

$$\hat{\varphi} = \arg\min_\varphi \sum_{t=1}^{T} D_{\text{KL}}\left(q(x_{t-1}^{\text{FD}} \mid x_t^{\text{FD}}, x^{\text{FD}}, x^{\text{LD}}) \,\|\, p_\varphi(x_{t-1}^{\text{FD}} \mid x_t^{\text{FD}}, x^{\text{LD}})\right) \quad (6)$$

Under the Markov assumption $x_t \perp x_{1:t-2} \mid x_{t-1}$, the true posterior factorizes into two Gaussians equation:

$$q(x_t^{\text{FD}} \mid x_{t-1}^{\text{FD}}, x^{\text{LD}}) = \mathcal{N}(x_t^{\text{FD}}; x_{t-1}^{\text{FD}} + \alpha_t e_0, \gamma^2 \alpha_t I)$$

$$q(x_{t-1}^{\text{FD}} \mid x_t^{\text{FD}}, x^{\text{LD}}) = \mathcal{N}(x_{t-1}^{\text{FD}}; x^{\text{FD}} + \beta_{t-1} e_0, \gamma^2 \beta_{t-1} I) \quad (7)$$

Multiplying them yields a single Gaussian equation

$$q(x_{t-1}^{FD} \mid x_t^{FD}, x^{FD}, x^{LD}) = \mathcal{N}\left(x_{t-1}^{FD}; \underbrace{\frac{\beta_{t-1}}{\beta_t} x_t^{FD} + \frac{\alpha_t}{\beta_t} x^{FD}}_{\breve{\mu}_q}, \underbrace{\gamma^2 \frac{\alpha_t \beta_{t-1}}{\beta_t} I}_{\breve{\Sigma}_q}\right) \quad (8)$$

Assuming $\Sigma_\varphi(t) = \Sigma_q(t)$, the KL in (6) reduces to matching the means:

$$\hat{\varphi} = \arg\min_\varphi \sum_{t=1}^{T} \frac{\beta_t}{2\gamma^2 \alpha_t \beta_{t-1}} \|\mu_\varphi(x_t^{FD}, x^{LD}, t) - \mu_q\|_2^2 \quad (9)$$

We choose $\mu_\varphi$ to match the true posterior mean under our forward noise schedule. Concretely, we parameterize

$$\mu_\varphi(x_t^{FD}, x^{LD}, t) = \frac{\beta_{t-1}}{\beta_t} x_t^{FD} + \frac{\alpha_t}{\beta_t} g_\varphi(x_t^{FD}, x^{LD}, t) \quad (10)$$

where $\{\alpha_t, \beta_t\}$ are known scalar schedules and $g_\varphi$ is trained to predict the clean image contribution. Under the simplifying assumption that the learned and true covariances coincide, minimizing the sum of KL divergences between forward and reverse kernels reduces to the familiar mean-squared loss:

$$\hat{\varphi} = \arg\min_\varphi \sum_{t=1}^{T} \|g_\varphi(x_t^{FD}, x^{LD}, t) - x^{FD}\|_2^2 \quad (11)$$

In practice, this means we train $g_\varphi$ at each time step to "denoise" $x_t^{FD}$ back toward the true full-dose image, thereby learning an efficient reverse diffusion that recovers high-quality CT images in just a few steps.

**Modified loss function.**

The model is optimized with a composite objective that combines MSE, MAE (L1), LPIPS, and total TV losses. This formulation enables joint optimization for pixel-wise fidelity, structural sharpness, perceptual quality, and image smoothness. To preserve theoretical consistency with diffusion modeling, the $\ell_2$/MSE loss is retained as the primary component, which corresponds to minimizing the KL divergence between the true forward process and the learned reverse kernel. To further improve perceptual fidelity and structural sharpness, we augment the MSE loss with LPIPS $\ell_p$ loss (Zhang *et al* 2018c), $\ell_1$, and TV terms:

$$\mathcal{L}_\varphi = \lambda_2 \|g_\varphi(x_t^{FD}, x^{LD}, t) - x^{FD}\|_2^2 + \lambda_p \ell_p\big(g_\varphi(x_t^{FD}, x^{LD}, t), x^{FD}\big) + \lambda_1 \|g_\varphi(x_t^{FD}, x^{LD}, t) - x^{FD}\|_1 + \lambda_{TV} \text{TV}\big(g_\varphi(x_t^{FD}, x^{LD}, t)\big) \quad (12)$$

where, $\lambda_2$ controls the weight of the MSE term, $\lambda_1$ the L1 loss, $\lambda_p$ the perceptual LPIPS loss, and $\lambda_T V$ the total variation regularizer. The TV term is defined as:

$$\text{TV}(x) = \sum_{i,j} w_{i,j} \sqrt{(x_{i+1,j} - x_{i,j})^2 + (x_{i,j+1} - x_{i,j})^2 + \epsilon} \qquad (13)$$

While the formulation allows for spatially adaptive pixel-wise weighting, in this study we empirically set $w_{i,j} = 1$ for simplicity, which corresponds to the conventional isotropic TV regularization.

In our study, we set $\lambda_2 = 4.0, \lambda_p = 1.0, \lambda_{TV} = 0.1$ and $\lambda_1 = 1$, based on small-scale validation sweeps. The MSE component remains dominant to ensure the theoretical link to KL minimization, while the other terms act as auxiliary regularization to guide perceptual and structural reconstruction. Although a full sensitivity analysis was not performed, empirical observations confirmed that moderate variations in these weights did not materially affect the outcomes, suggesting that the method is reasonably robust to the exact coefficient settings.

2.3 Implementation details

We assessed the performance of our proposed method using a consistent test set comprising 15 scans (5 chest scans and 10 abdomen scans). In this study, we use four full-reference image quality metrics to comprehensively evaluate our model: peak signal-to-noise ratio (PSNR) quantifies overall reconstruction accuracy by comparing the maximum possible signal power to the power of the reconstruction error; structural similarity index measure (SSIM)(Wang *et al* 2004) assesses preservation of local luminance, contrast, and structural information, reflecting texture and pattern fidelity; Visual information fidelity (VIFp)(Sheikh and Bovik 2005) estimates the amount of visual information from the reference image that is retained in the reconstruction, correlating closely with human perceptual judgment; and learned perceptual image patch similarity (LPIPS)(Sheikh and Bovik 2006a) computes perceptual distance based on deep features from a pretrained network, emphasizing semantic and fine-detail quality(Zhang *et al* 2018c),(Ding *et al* 2020). GMSD is a full-reference image quality assessment metric that quantifies the perceptual difference between a test image and its reference by evaluating local gradient structures(Xue *et al* 2013). For consistency with prior studies, we report PSNR, SSIM, and VIFp when comparing against existing methods. In addition, we include PSNR, SSIM, VIFp, GMSD, and LPIPS in our internal ablation studies. Together, these metrics capture both pixel-level fidelity and perceptual quality, providing a comprehensive basis for evaluating LDCT denoising performance(Kang *et al* 2017),(Sheikh and Bovik 2006b, Chen *et al* 2017c),(Eulig *et al* 2024).

Noise scheduler. We employ a non-uniform geometric noise scheduler. As shown in Figure 2, we control the forward corruption by two hyperparameters: a global scaling factor $\gamma$ and a time-dependent variance schedule $\{\beta_t\}_{t=1}^T$. For the intermediate timesteps $t \in [2, T-1]$, we have $\beta_t = (T-1)(\frac{t-1}{T-1})^p$, where p is a hyperparameter controlling the growth rate. As illustrated in Figure 2, smaller values of $p$ result in higher noise levels in the generated images throughout the forward diffusion steps. Prior studies have shown that the product $\gamma\sqrt{\alpha_t}$ must remain small for a neural network to faithfully learn the diffusion trajectory(Ho *et al* 2020a), so we fix $\gamma\sqrt{\alpha_t} = 0.04$, $\beta_1 = \left(\frac{0.04}{\gamma}\right)^2$, and choose $\gamma = 2$ to satisfy the lower-bound $\beta_1 \to 0$. To ensure $\beta_T \to 1$, we set $\beta_T = 0.99$. Rather than a linear schedule, in line with recent diffusion models, we adopt the nonuniform geometric parameters as $\sqrt{\beta_t}$ (Safari *et al* 2025, Yue *et al* 2024). Unlike $p$, which modulates the rate at which noise accumulates, a larger $\gamma$ amplifies the overall noise magnitude at each step. Panels (c)–(k) in Figure 2 demonstrate how varying $p$ and $\gamma$ alters the noise intensity of the forward diffusion process at different timesteps $t$.

Training Configuration. The proposed RE-EDPM model was implemented in PyTorch (v2.6.0+cu124) and executed on an NVIDIA A100 80GB PCIe GPU. Training was performed using a batch size of 4 for 50 epochs, totaling 13,360 steps. Optimization was conducted using the Rectified Adam (RAdam) optimizer(Liu *et al* 2019), with a cosine annealing learning rate schedule(Loshchilov and Hutter 2016). A warm-up phase of 5,000 steps was applied to stabilize the early stages of training.

Inference and Validation. Validation was conducted at the end of every training epoch (≈267 steps), resulting in 50 validation evaluations over the entire training process. At each validation stage, the current model checkpoint was applied to the held-out validation set, and reconstructed denoised images were quantitatively assessed using PSNR, SSIM, and VIF. These metrics were logged to monitor training progress, and the checkpoint with the highest mean SSIM was selected for final testing. Higher PSNR and SSIM, together with higher VIF, indicate better image restoration performance. For completeness, we also report GMSD and LPIPS in the Supplementary Table S1.

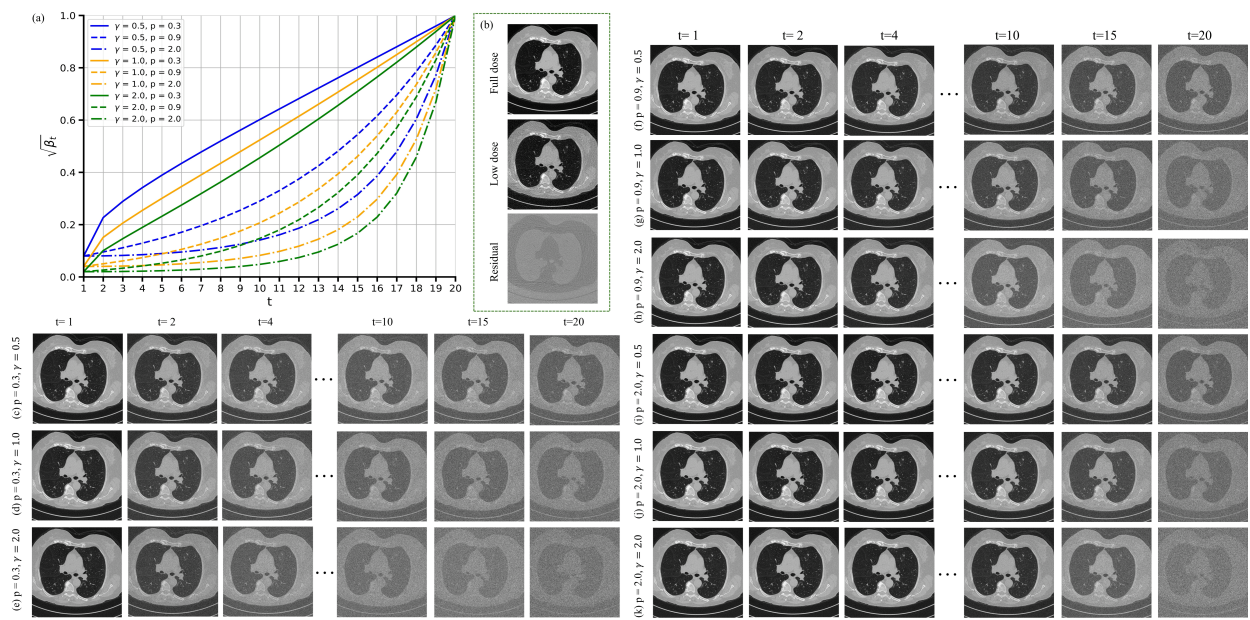

**Figure 2**: Schematic illustration of the influence of hyperparameters on the forward diffusion process. (a) shows the evolution of the noise scaling factor $\sqrt{\beta_t}$ as a function of the diffusion time step t for various hyperparameter combinations; (b) shows full dose CT image $x^{FD}$ as the ground truth, low dose CT image $x^{LD}$, and the residual error $e_0 = x^{LD} - x^{FD}$; Panels (c)-(e),(f)-(h) and (i)-(k) demonstrates the effect of varying the hyperparameter γ on the noise level of the generated images $x_t^{FD}$ across different values of $p$ with higher γ values produce stronger noise. (f) and (i) specifically compare the effect of p for a fixed γ.

## 3. Results

As shown in Table 1, our proposed method significantly outperformed previous state-of-the-art (SOTA) denoising approaches across both chest (10% dose) and abdomen (25% dose) CT imaging tasks. Specifically, it achieved the highest SSIM scores (0.879±0.068 for chest and 0.971±0.000 for abdomen), highest PSNR values (31.597±2.520 dB and 36.685±2.535 dB, respectively), and strongest perceptual quality indicated by the highest VIFP scores among all models (0.366±0.121 for chest and 0.510±0.007 for abdomen). These results indicate superior structural preservation, noise suppression, and perceptual fidelity.

Figure 3 summarizes both quantitative and qualitative results. Representative CT slices from the lowest- and highest-SSIM cases demonstrate that RE-EDPM more faithfully recovers fine anatomical details and tissue textures compared with competing methods, while difference maps confirm reduced residual errors relative to full-dose images. Additional qualitative comparisons with baseline methods are provided in Supplementary Figure S2.

Quantitative evaluation and qualitative inspection confirm that our approach outperforms classical denoising algorithms(Feruglio *et al* 2010), state-of-the-art CNN-based frameworks(Chen *et al* 2017b, 2017c) and GAN-driven architectures(Kwon and Ye 2021, Huang *et al* 2021, Isola *et al* 2017), and as well as the recent TransCT transformer (Zhang *et al* 2021) and Bilateral method(Wagner *et al* 2022), striking an optimal compromise between noise suppression and preservation of fine structural details.

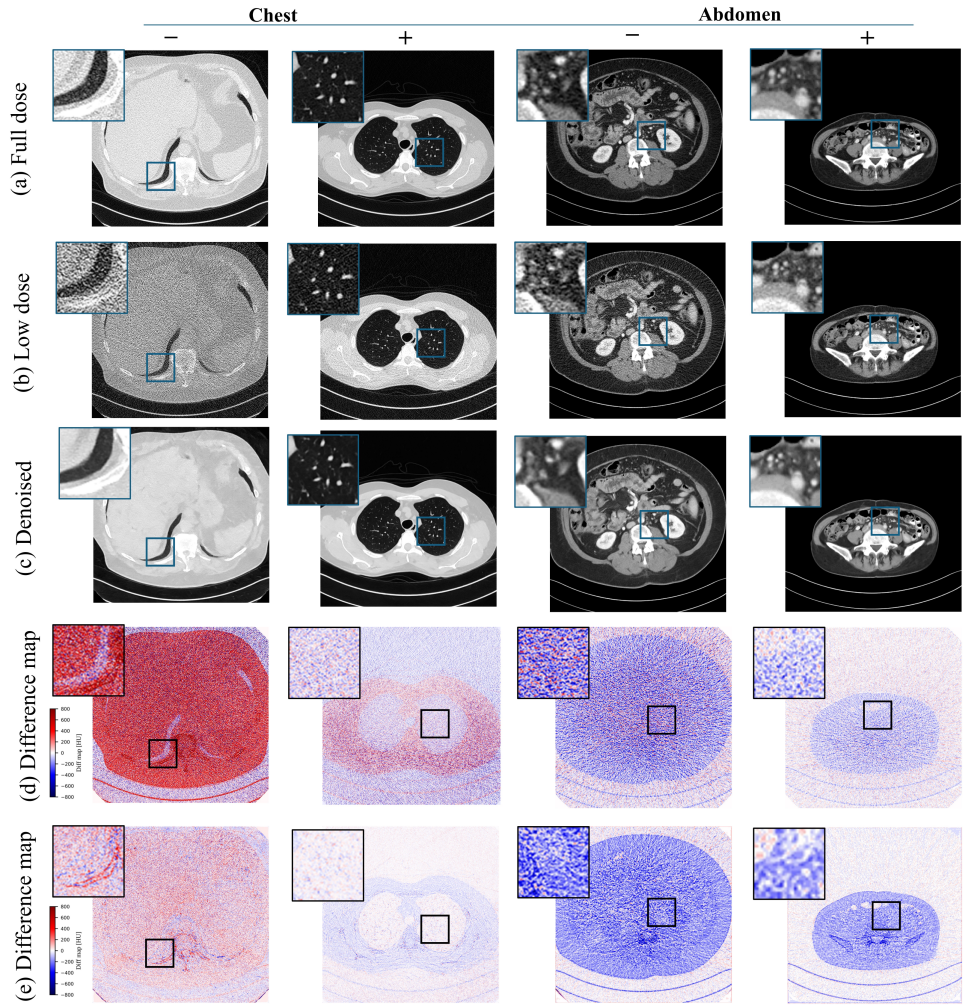

**Figure 3**: Representative CT slices denoised by the proposed RE-EDPM model. (a–c) Example slices corresponding to cases with the lowest (–) and highest (+) average SSIM achieved by RE-EDPM on the test dataset. (d) Quantitative difference map (in Hounsfield Units) between the low-dose and full-dose (ground truth) CT images. (e) Difference map (HU scale) between the RE-EDPM output and the FDCT image. CT slices for chest regions are displayed using a lung window (window center: –600 HU, window width: 1500 HU), while slices for abdominal regions are displayed using a soft tissue window (window center: +50 HU, window width: 400 HU). For qualitative comparisons with baseline methods, refer to Supplementary Figure S2.

Standard CNN denoisers can run in real time typically under 0.06 s per 512×512 slice on a GPU (equating to <20 s for a 300-slice volume), but often yield lower denoising fidelity than diffusion methods(Kang *et al* 2017, Chen *et al* 2017a, Zhang *et al* 2018a). Accelerated diffusion variants like DDIM achieve 10–50× faster sampling than vanilla DDPMs—bringing per-slice inference down from ~20 s to roughly 0.4–2 s—but still lag in quality or require trade-offs in step count(Ho *et al* 2020a). Based on our testing and previous studies, GAN-based frameworks report inference times of 0.5–3.8 s for an entire 360-slice volume, yet they can introduce over-smoothing or artifact issues(Kwon and Ye 2021, Huang *et al* 2021, Isola *et al* 2017). In contrast, by using just four diffusion steps, RE-EDPM denoises two slices in 0.25 s (≈0.125 s per slice), processing a full 300-slice CT in under 40 s orders of magnitude faster than DDPMs and within a clinically viable window, while still achieving superior PSNR, SSIM, VIFp, and LPIPS performance. Using the same U-Net backbone and computational precision, our method requires only 4 forward passes, compared with 10 for Fast-DDPM and 1000 for DDPM/DDIM. Despite this markedly smaller computational budget, the proposed 4-step sampler consistently achieves higher PSNR and SSIM values and lower LPIPS scores than Fast-DDPM at 10 steps and DDPM/DDIM at 1000 steps on the CT test set. Since the per-step computational cost is approximately constant under a fixed architecture, this corresponds to an effective reduction in inference cost by a factor of about 2.5 relative to Fast-DDPM and about 250 relative to DDPM, while maintaining or even improving reconstruction quality. For clarity, we report computational efficiency in terms of the number of function evaluations rather than wall-clock latency, as the latter depends on implementation and hardware. Compared with other approaches, RE-EDPM strikes a unique balance between speed and quality. We summarize the hyper-parameter search ranges and the final selected values for each method, including classical approaches (BM3D, Bilateral filtering), CNN-based models (CNN-10, RED-CNN, ResNet, QAE), GAN-based models (WGAN-VGG, CycleGAN, Pix2Pix, DU-GAN), Transformer-based models (TransCT), and diffusion-based models (DDPM, Fast-DDPM, DDIM, and our proposed method). Details of implementation are provided in Supplementary **Tables S1**.

**Table 1.** Quantitative comparison of denoising models on Chest (10% dose) and Abdomen (25% dose) CT datasets.

|  | Chest(10% dose) | | | Abdomen (25% dose) | | |
| --- | --- | --- | --- | --- | --- | --- |
| Models | SSIM[-]↑ | PSNR[dB]↑ | VIFp[-]↑ | SSIM[-]↑ | PSNR[dB]↑ | VIFp[-]↑ |
| Proposed method | **0.879±0.068** | **31.597±2.520** | **0.366±0.121** | **0.971±0.000** | **36.685±2.535** | **0.510±0.007** |
| Fast-DDPM(Jiang *et al* 2025) | 0.843±0.117 | 30.132±2.721 | 0.310±0.126 | 0.912±0.403 | 34.122±2.688 | 0.472±0.096 |
| DDPM(Ho *et al* 2020b) | 0.737±0.117 | 29.342±2.280 | 0.270±0.076 | 0.812±0.081 | 33.442±2.674 | 0.412±0.096 |
| DDIM(Song *et al* 2020a) | 0.710±0.134 | 29.144±2.120 | 0.320±0.083 | 0.812±0.081 | 32.232±2.894 | 0.386±0.114 |
| Cycle-GAN(Kwon and Ye 2021) | 0.726±0.134 | 26.763±4.160 | 0.246±0.087 | 0.942±0.093 | 32.306±3.115 | 0.454±0.102 |
| Pixel2pixel-GAN(Isola *et al* 2017) | 0.738±0.122 | 30.742±2.592 | 0.342±0.128 | 0.804±0.074 | 33.494±2.119 | 0.395±0.099 |
| BM3D(Alnuaimy *et al* 2024) | 0.553±0.132 | 26.340±2.340 | 0.172±0.002 | 0.868±0.034 | 30.500±0.200 | 0.372±0.117 |
| CNN-10(Chen *et al* 2017c) | 0.587±0.001 | 27.710±0.020 | 0.192±0.001 | 0.896±0.001 | 32.400±0.100 | 0.449±0.003 |
| RED-CNN(Chen *et al* 2017b) | 0.609±0.002 | 28.360±0.030 | 0.221±0.003 | 0.903±0.001 | 33.220±0.070 | 0.491±0.008 |
| WGAN-VGG(Yang *et al* 2018) | 0.510±0.030 | 25.500±0.200 | 0.148±0.004 | 0.882±0.002 | 30.500±0.900 | 0.380±0.010 |
| ResNet(Missert *et al* 2018) | 0.610±0.001 | 28.420±0.030 | 0.224±0.002 | 0.901±0.002 | 33.150±0.080 | 0.487±0.006 |
| QAE(Fan *et al* 2019) | 0.584±0.003 | 27.620±0.090 | 0.186±0.003 | 0.894±0.002 | 32.000±0.200 | 0.418±0.007 |
| DU-GAN(Huang *et al* 2021) | 0.565±0.004 | 26.700±0.100 | 0.168±0.002 | 0.894±0.002 | 32.100±0.300 | 0.427±0.005 |
| TransCT(Zhang et al 2021) | 0.563±0.002 | 26.990±0.050 | 0.167±0.002 | 0.877±0.003 | 30.500±0.200 | 0.372±0.007 |
| Bilateral(Wagner et al 2022) | 0.555±0.001 | 25.590±0.040 | 0.159±0.002 | 0.859±0.003 | 27.100±0.100 | 0.361±0.003 |

**Ablation study on loss function design**

To evaluate the impact of different loss functions on model performance, we conducted an ablation study using five combinations of commonly used objective terms: MAE, LPIPS, and TV. In all ablation settings, the $\ell_2$/MSE loss was always included as the dominant component by default; for simplicity, it is not explicitly listed in Table 2. Results for both chest (10% dose) and abdomen (25% dose) CT images are summarized in Table 2. Across both anatomical regions, the combination of TV + LPIPS + MAE consistently achieved the best overall performance. For the chest dataset, it yielded the highest SSIM (0.879±0.068) and VIFP (0.366±0.121). Similarly, on the abdomen dataset, it achieved the highest SSIM (0.971±0.000) and competitive scores across all metrics, with PSNR of 36.685±2.535 dB and VIFP of 0.510±0.007. While MAE alone achieved slightly higher SSIM for the chest region (0.892±0.062), it introduced a significant increase in LPIPS (0.344±0.105), suggesting a trade-off between pixel-wise accuracy and perceptual quality. In contrast, LPIPS alone produced competitive perceptual scores but failed to capture fine structural information, as indicated by its lower VIFP and PSNR. The inclusion of TV loss appeared to enhance structural smoothness and detail preservation, as reflected in improved perceptual and structural scores for combinations involving TV (e.g., TV+LPIPS, TV+MAE). However, only when all three components (TV + LPIPS + MAE) were jointly applied did the model achieve both perceptual and quantitative robustness.

These findings confirm that the composite loss function (MSE + TV + LPIPS + MAE) offers the best trade-off among pixel-level fidelity, perceptual similarity, and structural preservation, and is thus adopted in our final RE-EDPM model. For consistency with the main manuscript, PSNR, SSIM, and VIF are reported as the primary evaluation metrics in Table 2, while LPIPS and GMSD are provided only in Supplementary Table S1 to further highlight perceptual and structural quality. To further validate the superiority of the TV+LPIPS+MAE composite loss, we conducted pairwise statistical comparisons using the Wilcoxon signed-rank test across all loss combinations. The analysis was performed separately for PSNR, NMSE, SSIM, VIFp, and GMSD metrics, with Bonferroni correction applied to account for multiple comparisons. For the chest CT dataset, TV+LPIPS+MAE consistently and significantly outperformed all other loss combinations ($p < 0.05$ after correction) across most metrics. Specifically, compared to the next-best models (e.g., LPIP_TV and MAE_TV), the proposed loss achieved highly significant improvements in PSNR (corrected $p = 4.89\text{e-}12$ vs. LPIP_TV, $p = 6.18\text{e-}84$ vs. MAE), NMSE (corrected $p = 7.27\text{e-}09$ vs. LPIP_TV,

p = 4.79e-92 vs. MAE), SSIM (corrected p = 9.33e-03 vs. LPIP_TV, p = 9.14e-251 vs. MAE), and VIFp (corrected p = 5.06e-262 vs. LPIP_TV, p = 5.32e-120 vs. MAE). Similar trends were observed for GMSD, where TV+LPIPS+MAE demonstrated the lowest error, with highly significant p-values (e.g., p = 3.31e-77 vs. LPIP_TV, p = 8.37e-234 vs. MAE).

These statistical findings further reinforce that the proposed composite loss not only yields the best perceptual and quantitative performance but also does so with strong statistical confidence. Thus, integrating TV, LPIPS, and MAE within the RE-EDPM framework provides a significant and robust advantage for LDCT image denoising. On both the chest and abdominal CT datasets, our proposed TV + LPIPS + MAE composite loss yielded statistically significant improvements over every other loss combination (all pairwise comparisons p < 0.05), achieving the best balance of noise suppression and structural fidelity under the RE-EDPM framework.

**Table 2.** Ablation study on the effect of loss functions for chest (10% dose) and abdomen (25% dose).

| Models | chest (10% dose) | | | Abdomen (25% dose) | | |
| --- | --- | --- | --- | --- | --- | --- |
| | SSIM[-]↑ | PSNR[dB]↑ | VIFp[-]↑ | SSIM[-]↑ | PSNR[dB]↑ | VIFp[-]↑ |
| LPIPS | 0.877±0.068 | 31.154±2.589 | 0.353±0.109 | 0.960±0.000 | 34.583±2.166 | 0.496±0.007 |
| MAE | 0.892±0.062 | 30.545±2.682 | 0.341±0.109 | 0.970±0.000 | 36.551±2.571 | 0.512±0.007 |
| LPIPS+MAE | 0.878±0.068 | 31.104±2.535 | 0.333±0.124 | 0.964±0.000 | 36.324±2.329 | 0.509±0.008 |
| TV+MAE | **0.893±0.059** | 31.287±2.623 | 0.312±0.099 | 0.965±0.000 | 36.614±2.562 | 0.505±0.007 |
| TV+LPIPS | 0.879±0.067 | 31.157±2.56 | 0.333±0.113 | 0.963±0.000 | 35.996±2.369 | 0.502±0.008 |
| TV+LPIPS+MAE | 0.879±0.068 | **31.597±2.52** | **0.366±0.121** | **0.971±0.000** | **36.685±2.535** | 0.510±0.007 |

## 4. Discussion

Computed tomography remains a key imaging modality in clinical diagnostics, valued for its rapid acquisition and anatomical clarity. However, concerns over radiation exposure have led to growing interest in LDCT. While dose reduction mitigates radiation risk, it inevitably introduces significant noise and artifacts, which degrade image quality and diagnostic reliability. To address this trade-off, we propose RE-EDPM, an efficient denoising framework based on probabilistic diffusion modeling. Unlike conventional diffusion models that require dozens of iterative sampling steps, RE-EDPM introduces a residual-aware mechanism that integrates the initial error $e_0$ between low-dose and full-dose images directly into the forward diffusion process. By aligning the forward and reverse diffusion processes, our model more efficiently learns the inverse mapping, requiring only a handful of sampling steps to denoise CT images and thereby dramatically reducing inference time compared to standard diffusion-based approaches.

Experiments on both chest and abdominal LDCT datasets demonstrate that RE-EDPM significantly enhances both visual quality and computational efficiency. In particular, it outperforms baseline methods in restoring fine anatomical structures with reduced residual artifacts. Quantitative evaluations consistently show that our model achieves higher PSNR and SSIM, along with lower LPIPS and GMSD values, indicating more accurate and perceptually faithful denoising results. The reduced variance across test cases also suggests enhanced model stability, attributed to the residual-guided diffusion formulation.

The RE-EDPM method has the potential to extend beyond CT denoising and be adapted to other low-quality-to-high-quality image translation tasks. Further enhancements, such as adaptive noise scheduling or hybrid perceptual loss functions, may yield additional improvements in restoration performance and generalization capability. A key advantage of our approach is its greatly reduced inference latency compared to conventional DDPMs. Whereas a vanilla DDPM often requires on the order of 1,000 forward-backward diffusion steps per slice translating to roughly 15–30 seconds of GPU time for a single CT slice. Our model completes denoising in only 0.25 s for a batch of two slices on an NVIDIA A6000 for inference. In practical terms, for a typical thoracic CT exam of ∼300 slices, we can process the entire volume in under 40 s, versus more than 90 min for a standard DDPM, an improvement of over two orders of magnitude. This speedup makes real-time or near real-time clinical deployment feasible, particularly in busy radiology workflows where scan turnaround is critical. At inference, our residual-shifting sampler requires only four model forward passes (T = 4), whereas typical DDPM/DDIM samplers use hundreds to a thousand with the same backbone. For a fixed architecture and precision, the per-pass cost is approximately constant, so the inference compute budget scales with the number of forward passes (NFE). We therefore claim a reduction in compute rather than wall-clock latency, which depends on hardware and implementation and is not evaluated here. Training is performed with standard continuous-time sampling and is decoupled from the inference step budget. Bridging from the deterministic endpoint $x_T = y$ induces a stiff reverse field around a delta distribution; with a small step budget, discretization errors are amplified. Drawing $x_T \sim \mathcal{N}(y, \gamma^2 I)$ regularizes the initial condition, makes the reverse dynamics well-posed, and allows controlled exploration before contraction. The residual shift in the mean aligns the forward trajectory with the observation, shortening the transport path, while the non-zero covariance prevents early over-commitment to artifacts present in $y$. Ablations against DDPM/DDIM/Fast-DDPM (which omit residual shifting) corroborate that the noisy initialization plus residual guidance is the main driver behind our 4-step efficiency and robustness. RE-EDPM's rapid inference (only four sampling steps per slice) and robust restoration quality make it a promising candidate for real-time clinical workflows. Empirically, this design yields stable reconstructions with only four reverse updates, whereas standard DDPM, DDIM, and Fast-DDPM baselines (which do not use residual shifting) require substantially more steps to reach comparable quality. Future work will explore adaptive noise scheduling, task-specific perceptual losses, and extensions to other imaging modalities (e.g.,

Cone Beam CT or PET CT), aiming to generalize the framework for broader low-quality–to–high-quality image translation tasks in medical imaging.

This study has several limitations that warrant further investigation. First, although RE-EDPM processes 2D CT slices independently, leveraging the efficiency of our Swin-UNet backbone, it does not explicitly enforce volumetric consistency across adjacent slices, which may lead to subtle inter-slice discontinuities in clinical reconstructions. Second, our residual-shifting mechanism and noise schedule are optimized for the specific 10 % and 25 % dose settings used during training; performance may degrade when applied to scans with different dose reductions or noise characteristics without additional domain adaptation. Third, by restricting inference to just four diffusion steps, we achieve exceptional speed but may sacrifice some fine-scale detail in highly textured regions, such as small vessels or subtle ground-glass opacities. While our hybrid regularization loss enhances perceptual fidelity, RE-EDPM does not provide explicit uncertainty quantification for each voxel, limiting its transparency in edge-case scenarios. Future work will explore 3D or slice-aware extensions to enforce volumetric smoothness, adaptive noise scheduling to accommodate a broader range of dose levels, and integrated uncertainty estimation to improve robustness and clinical interpretability. A remaining limitation is that our current implementation denoises axial slices independently, which can induce subtle inter-slice flicker visible in coronal and sagittal reconstructions. To promote 3D coherence, we will (i) adopt 2.5D conditioning that supplies each slice with shallow context from its neighbors, (ii) standardize intensities at the volume level and use overlapping inference along the z-axis with weighted blending, and (iii) investigate memory-efficient 3D backbones/diffusion with sliding-window inference to learn volumetric priors end-to-end. When projection data are available, we will incorporate physics-based data-fidelity updates that couple slices through the acquisition geometry and add lightweight volumetric-consistency regularization. Effectiveness will be assessed using 3D PSNR and SSIM, z-continuity metrics, and performance on downstream 3D tasks.

In summary, RE-EDPM integrates both perceptual- and structure-aware loss components and leverages a lightweight residual-guided diffusion process to achieve high-fidelity CT image restoration in significantly fewer steps. This makes it a promising framework for accelerating LDCT workflows while preserving diagnostic quality.

## 5. Conclusion

We have introduced RE-EDPM for LDCT denoising that integrates a residual-shifting mechanism into the forward diffusion process and employs a lightweight Swin-based U-Net backbone to align LDCT and full-dose distributions while capturing long-range dependencies. A hybrid regularization loss combining pixel-level reconstruction, perceptual similarity, and spatially adaptive TV—suppresses noise without sacrificing fine anatomical detail. Extensive evaluations on chest (10% dose) and abdomen (25% dose) datasets show that RE-EDPM outperforms both classical and recent transformer-based methods across PSNR, SSIM, VIFp, and LPIPS, with residual and difference maps confirming superior structural fidelity and reduced artifacts. Ablation studies validate the individual contributions of the residual-shifting mechanism and composite loss, and pairwise statistical tests establish the significance of these gains. Crucially, by using only four diffusion steps, our model denoises are much faster than a standard DDPM, making real-time clinical deployment feasible.

**Data availability**

The low-dose CT and projection datasets analyzed during this study are publicly accessible via The Cancer Imaging Archive at the following URL:

https://www.cancerimagingarchive.net/collection/ldct-and-projection-data/

**Acknowledgments**

This research is supported in part by the National Institutes of Health under Award Numbers R01CA272991, R01EB032680, P30CA008748 and U54CA274513.

**Conflict of Interest**

The authors have declared no conflicts of interest.

Safari M, Wang S, Eidex Z, Li Q, Middlebrooks E H, Yu D S and Yang X 2025 MRI super-resolution reconstruction using efficient diffusion probabilistic model with residual shifting *ArXiv Prepr. ArXiv250301576*

Safari M, Yang X and Fatemi A 2024 MRI data consistency guided conditional diffusion probabilistic model for MR imaging acceleration *Medical Imaging 2024: Clinical and Biomedical Imaging* vol 12930 (SPIE) pp 202–5

Sheikh H R and Bovik A C 2005 A visual information fidelity approach to video quality assessment *The first international workshop on video processing and quality metrics for consumer electronics* vol 7 (sn) pp 2117–28

Sheikh H R and Bovik A C 2006a Image information and visual quality *IEEE Trans. Image Process.* **15** 430–44

Sheikh H R and Bovik A C 2006b Image information and visual quality *IEEE Trans. Image Process.* **15** 430–44

Shu Z and Entezari A 2024 RBP-DIP: Residual back projection with deep image prior for ill-posed CT reconstruction *Neural Netw.* **180** 106740

Sidky E Y, Kao C-M and Pan X 2006 Accurate image reconstruction from few-views and limited-angle data in divergent-beam CT *J. X-Ray Sci. Technol.* **14** 119–39

Sidky E Y and Pan X 2008 Image reconstruction in circular cone-beam computed tomography by constrained, total-variation minimization *Phys. Med. Biol.* **53** 4777

Singh P, Diwakar M, Gupta R, Kumar S, Chakraborty A, Bajal E, Jindal M, Shetty D K, Sharma J, Dayal H, and others 2022 A method noise-based convolutional neural network technique for CT image denoising *Electronics* **11** 3535

Smith-Bindman R, Chu P W, Azman Firdaus H, Stewart C, Malekhedayat M, Alber S, Bolch W E, Mahendra M, Berrington de González A and Miglioretti D L 2025 Projected Lifetime Cancer Risks From Current Computed Tomography Imaging *JAMA Intern. Med.* Online: https://doi.org/10.1001/jamainternmed.2025.0505

Song J, Meng C and Ermon S 2020a Denoising diffusion implicit models *ArXiv Prepr. ArXiv201002502*

Song Y, Sohl-Dickstein J, Kingma D P, Kumar A, Ermon S and Poole B 2020b Score-based generative modeling through stochastic differential equations *ArXiv Prepr. ArXiv201113456*

Su B, Dong P, Hu X, Wang B, Zha Y, Wu Z and Wan J 2025 Fast and detail-preserving low-dose CT denoising with diffusion model *Biomed. Signal Process. Control* **105** 107580

Wagner F, Thies M, Gu M, Huang Y, Pechmann S, Patwari M, Ploner S, Aust O, Uderhardt S, Schett G, and others 2022 Ultralow-parameter denoising: trainable bilateral filter layers in computed tomography *Med. Phys.* **49** 5107–20

Supplement

Table S1. Quantitative results of the ablation study on loss function design with additional perceptual and structural quality metrics.

| Models | Chest (10% dose) | | | | | Abdomen (25% dose) | | | | |
|---|---|---|---|---|---|---|---|---|---|---|
| | SSIM[-]↑ | PSNR[dB]↑ | VIFp[-]↑ | GMSD[-]↓ | LPIPS[-]↓ | SSIM[-]↑ | PSNR[dB]↑ | VIFp[-]↑ | GMSD[-]↓ | LPIPS[-]↓ |
| LPIPS | 0.877±0.068 | 31.154±2.589 | 0.353±0.109 | 0.042±0.014 | 0.082±0.041 | 0.960±0.000 | 34.583±2.166 | 0.496±0.007 | 0.014±0.000 | 0.043±0.000 |
| MAE | 0.892±0.062 | 30.545±2.682 | 0.341±0.109 | 0.048±0.017 | 0.344±0.105 | 0.970±0.000 | 36.551±2.571 | 0.512±0.007 | 0.014±0.000 | 0.038±0.019 |
| LPIPS+MAE | 0.878±0.068 | 31.104±2.535 | 0.333±0.124 | 0.042±0.014 | 0.083±0.041 | 0.964±0.000 | 36.324±2.329 | 0.509±0.008 | 0.013±0.000 | 0.041±0.000 |
| TV+MAE | **0.893±0.059** | 31.287±2.623 | 0.312±0.099 | 0.05±0.017 | 0.097±0.062 | 0.965±0.000 | 36.614±2.562 | 0.505±0.007 | 0.013±0.000 | 0.041±0.000 |
| TV+LPIPS | 0.879±0.067 | 31.157±2.56 | 0.333±0.113 | 0.042±0.014 | 0.083±0.041 | 0.963±0.000 | 35.996±2.369 | 0.502±0.008 | 0.014±0.000 | 0.042±0.000 |
| TV+LPIPS+MAE | 0.879±0.068 | **31.597±2.52** | **0.366±0.121** | 0.042±0.014 | 0.084±0.042 | **0.971±0.000** | **36.685±2.535** | 0.510±0.007 | 0.014±0.000 | 0.038±0.019 |

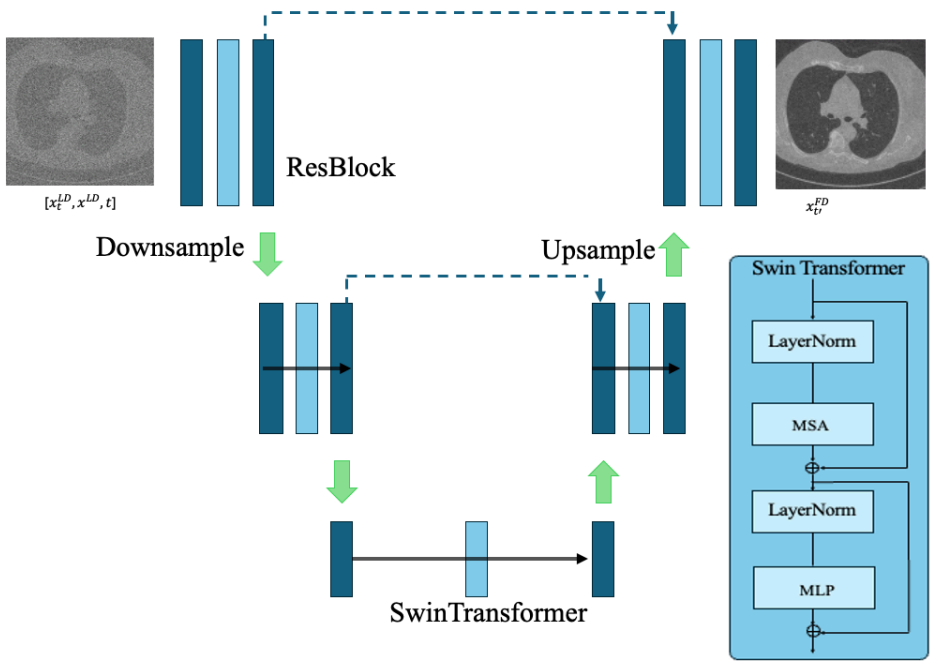

Figure S1. Illustration of the diffusion U-Net architecture of the proposed RE-EDPM.
The model is adapted from the widely used diffusion U-Net, with conventional attention layers replaced by Swin Transformer blocks. Each Swin Transformer block consists of LayerNorm, multi-head self-attention (MSA), and an MLP, enabling the network to better handle images of varying resolutions and enhance structural detail preservation. The inputs consist of a LDCT image $x^{LD}$, a diffused low-dose image $x_t^{LD}$ at a given diffusion step $t$, and the corresponding timestep information. The output is the estimated denoised image $x_{t'}^{FD}$ that approximates the FDCT target $x^{FD}$, where $t' < t$.

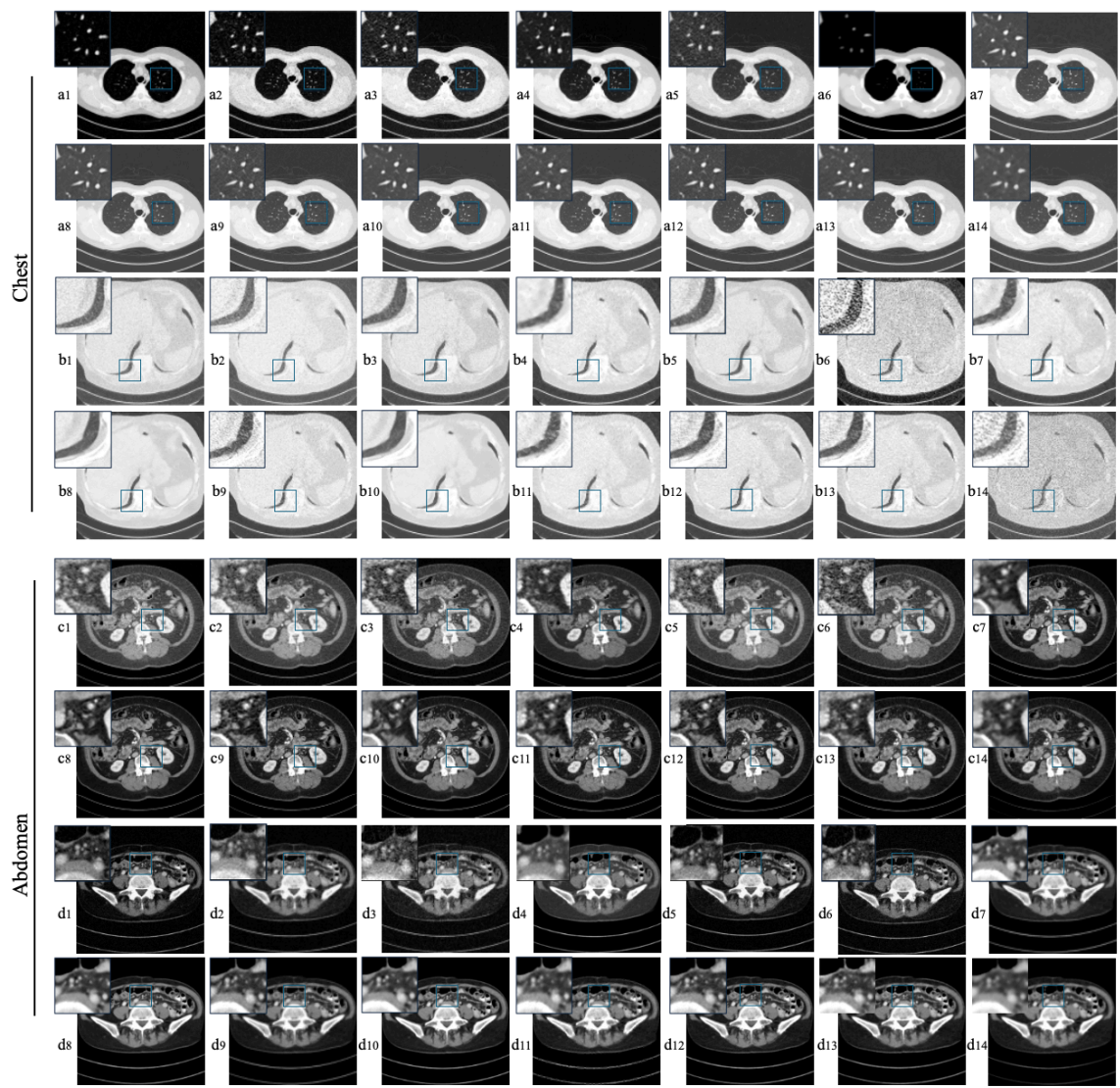

Figure S2. Representative qualitative comparison across baseline methods for LDCT denoising. (a1-a14) correspond to methods 1--14: 1. FastDDPM, 2. DDPM, 3. DDIM, 4. CycleGAN, 5. Pixel2pixel, 6. BM3D, 7. CNN-10, 8. RED-CNN, 9. WGAN-VGG, 10. ResNet, 11. QAE, 12. DU-GAN, 13. TransCT, 14. Bilateral.